\documentstyle[aps,prb,twocolumn,epsfig]{revtex}

\newcommand{\beq}{\begin{equation}}
\newcommand{\eeq}{\end{equation}}
\newcommand{\bea}{\begin{eqnarray}}
\newcommand{\eea}{\end{eqnarray}}

\begin{document}

\title{Frustration of antiferromagnetism in the \\ $t-t'$-Hubbard model at weak
coupling}

\author{W. Hofstetter and D. Vollhardt}
\address{Theoretische Physik III, Elektronische Korrelationen 
und Magnetismus, Universit\"at Augsburg, D-86135 Augsburg, Germany}

\date{February 19, 1998}

\maketitle

\begin{abstract}
The perfect-nesting instability towards antiferromagnetism of the 
Hubbard model is suppressed by next-nearest 
neighbor hopping $t'$. The asymptotic behavior
of the critical coupling $U_c(t')$ at small $t'$ is calculated in
dimensions $d=2,3,\infty$ 
using Hartree theory; this yields the exact result at least in $d>2$.
The order of the transition is also determined.
A region of stability of a metallic antiferromagnetic phase in $d=3$
is identified.
\end{abstract}

\section{Introduction}
Microscopic investigations of strongly correlated electron systems are
usually based on lattice models, the simplest of which is the one-band
Hubbard model. In this model two particularly important features of real
materials, the structure and the dimension of the underlying lattice,
enter only via the kinetic energy. In the past, most model calculations
started from a given lattice (mostly hypercubic) and, for simplicity,
assumed the hopping to be restricted to nearest neighbors (NN). In this
case lattices are either bipartite or not. In dimensions $d>1$ and at half
filling bipartite lattices generically lead to an antiferromagnetic
instability in the ground state for any $U>0$ due to Fermi surface
nesting.

It has recently become clear, however, that pure NN-hopping with
amplitude $t$ cannot account for some important features of correlated
systems and that next-nearest neighbor (NNN) hopping with amplitude $t'$
has to be included for both qualitative and quantitative reasons.
Indeed, realistic band structures, photoemission data and
neutron-scattering measurements of high-${\rm T_c}$ and related materials
usually cannot be fit by an energy dispersion that originates solely
from NN-hopping \cite{Benard 93,Tohyama 93,Stemmann 94,Andersen 95};
estimates for $|t'/t|$ range from 0.15 to 0.5.
A particularly clear-cut example is the antiferromagnetic insulator
${\rm Sr_2 CuO_2 Cl_2}$ where the inclusion of a $t'$-term is essential to
understand the ARPES data \cite{Andersen 95}. 
The hole motion in the $t$-$J$-model is
also greatly affected by NNN-hopping \cite{Schiller 95}. 
Furthermore, it became
clear only most recently that the stability of metallic ferromagnetism
in the one-band Hubbard model strongly depends on the presence of a
$t'$-term in all dimensions $d\ge 1$ 
\cite{Tasaki 95,Penc 96,Daul 97,Hlubina 97,Ulmke 98,Wahle 97}. 

On bipartite lattices in $d\ge 2$ the $t'$-term has profound consequences: 
the frustration introduced by $t'$ suppresses the perfect-nesting instability
towards antiferromagnetism at small $U$. Up to now this effect was mainly
studied in $d=2$. Lin and Hirsch \cite{Lin 87} calculated magnetic properties
within Hartree theory and by exact diagonalization of small systems.
Tremblay et al.\ \cite{Benard 93} studied both magnetic 
and pairing correlations
by quantum Monte-Carlo simulations and a two-particle self-consistent
approach. Duffy and Moreo \cite{Duffy 95} investigated the renormalization of
$t'$ by the interaction $U$, and presented results \cite{Duffy 97} 
suggesting the
existence of an itinerant antiferromagnetic ground state at
half-filling. Metallic antiferromagnetic phases are known to exist, for
example, in ${\rm V_{2-x} O_3}$ \cite{Carter 91} and ${\rm NiS_{2-x} Se_x}$
\cite{Sudo 92}.

In this paper we wish to obtain analytic insight into the suppression of
the antiferromagnetic nesting instability at small $U$ and half filling
due to the frustration introduced by $t'$. To this end we solve the 
$t$-$t'$ Hubbard model 
\begin{eqnarray} \label{Hamiltonian}
{\cal H} &=& -\sum_{<ij> \atop \sigma} t\;(c_{i\sigma}^{\dagger} 
c_{j\sigma}^{\phantom{\dagger}} 
+ h.c.) + \sum_{(ij)\atop  \sigma} t'\;(c_{i\sigma}^\dagger\;
c_{j\sigma}^{\phantom{\dagger}} + h.c.)  
\nonumber  \\ 
&& +\, U\;\sum_{i} (n_{i \uparrow} - \frac{1}{2})(n_{i\downarrow} - \frac{1}{2}) - 
\mu\sum_{i\sigma} n_{i\sigma}.
\end{eqnarray}
within Hartree theory at $T=0$ and thereby calculate $U_c(t')$, the
critical value of $U$ for the onset of antiferromagnetism. For small $t'$
the results for $U_c$ become asymptotically exact\cite{van Dongen statement}, 
at least in dimensions $d>2$. 

We will calculate $U_c(t')$ explicitly in $d=2,3,\infty$, both numerically 
and analytically and will also determine the order of the transition into
the antiferromagnetic state.

\section{Hartree equations}
The Hubbard interaction is purely local and therefore does not include any
direct exchange. Consequently, there is no Fock term and Hartree-Fock theory
reduces to Hartree theory.
Within that approach 
spontaneous symmetry breaking with  
an antiferromagnetic order parameter (N\'eel order) is introduced by the \emph{ansatz}
\mbox{$\langle n_{i\sigma} \rangle = \frac{1}{2} (1+\sigma (-1)^{|i|} \Delta )$} with 
$\sigma=\pm 1$,  
where $\Delta$ is the staggered magnetization. We can then decompose the electron 
number operator into its expectation value and a fluctuating part $:n_{i\sigma}:$
according to $ n_{i\sigma} = :n_{i\sigma}: + \langle n_{i\sigma} \rangle $.
Neglecting terms quadratic in the fluctuations, the Hartree Hamiltonian takes the 
form
\bea
\lefteqn{{\cal H}_{H} = \frac{U N \Delta^2}{4}+}  \\
&& \sum_{\sigma \atop {\bf k} \in \frac{1}{2} B.Z.} 
\left(c^\dagger_{{\bf k}\sigma}, c^\dagger_{{\bf k}+{\bf Q}\,\sigma} \right) 
\left(\begin{array}{cc} \epsilon_{\bf k}  - \mu& -\sigma \frac{U\Delta}{2} \\
-\sigma \frac{U\Delta}{2} & \epsilon_{{\bf k} + {\bf Q}}  - \mu
\end{array} \right) 
\left( \begin{array}{c} c_{{\bf k}\sigma}^{\phantom{\dagger}} 
\\ c_{{\bf k}+{\bf Q}\,\sigma}^{\phantom{\dagger}} 
\end{array} \right),  \nonumber 
\eea
where the summation is restricted to half of the Brillouin zone.
Here we have introduced the number of sites $N$ and the nesting
vector \mbox{${\bf Q} = (\pi,\ldots ,\pi)$}.
The dispersion of the free electrons on the lattice is given by 
\begin{equation}  \label{free_electrons}
\epsilon_{\bf k} = -2 t\sum_{i=1}^d \cos k_i + 4 t' \sum_{i<j}^d \cos k_i \cos k_j.
\end{equation}
For $t'\ne 0$ the perfect nesting property $\epsilon({\bf k}) = -\epsilon({\bf k}+
{\bf Q})$ of the free dispersion no longer holds.
In the following we shall use the notation ${\bf k}' = {\bf k} +{\bf Q}$.
After diagonalization of the above quadratic form the mean-field Hamiltonian can be
written as
\begin{equation}
{\cal H}_{H} = \sum_{{\bf k}\sigma} (\tilde{\epsilon}_{\bf k} - \mu) 
\tilde{c}^\dagger_{{\bf k}\sigma} \tilde{c}_{{\bf k}\sigma}^{\phantom{\dagger}} 
+ \frac{U N \Delta^2}{4} 
\end{equation}
with the single particle energies 
\begin{equation} \label{HFenergies}
\tilde{\epsilon}_{\bf k} = \frac{1}{2} \left(\epsilon_{\bf k} + \epsilon_{{\bf k}'} + 
(\epsilon_{\bf k} - \epsilon_{{\bf k}'}) \sqrt{1+ \frac{U^2 
\Delta^2}{(\epsilon_{\bf k} - 
\epsilon_{{\bf k}'})^2}}\right).
\end{equation}
Since $\epsilon_{\bf k}$ does not fulfill the perfect nesting condition,
one has to adjust the chemical potential as a function of $\beta$, $\Delta$, $U$ and $t'$
to ensure the correct filling $n=1$.
In addition, the value of the order parameter $\Delta$ is determined self-consistently
by the condition that the free energy 
\begin{equation}
F[\beta,\Delta,U] = -\beta^{-1} \ln \textrm{Tr} \exp(-\beta {\cal H})
+\mu N,
\end{equation}
be minimal with respect to $\Delta$.
This leads to two self-consistency equations:
\begin{eqnarray}
1 &=& \frac{2}{N} \int \frac{d^d k}{(2\pi)^d}  \textrm{f}(\beta, \tilde{\epsilon}_{\bf k} - \mu) \label{self2} \\
1 &=& \frac{2 U}{N} \int \frac{d^d k}{(2\pi)^d}
\textrm{f}(\beta, \tilde{\epsilon}_{\bf k} - \mu) 
\frac{1}{(\epsilon_{\bf k} - \epsilon_{{\bf k}'}) \sqrt{1 + \frac{U^2 \Delta^2}
{(\epsilon_{\bf k} - \epsilon_{{\bf k}'})^2}}}, \label{self1} 
\end{eqnarray}
where $\textrm{f}(\beta,\tilde{\epsilon}_{{\bf k}} - \mu)$ 
is the Fermi function. 
The trivial solution $\Delta=0$ of the minimum 
condition has been excluded in (\ref{self1}). In the following, (\ref{self1}) and 
(\ref{self2}) will be analyzed for dimensions $d=2,3$ and $\infty$ at $T=0$. 

\section{Results and discussion}

\subsection{$d=2$} \label{d_equals_two}
We set $t=1$.
The presence of $t'\ne 0$ prevents an analytic evaluation
of the integrals in (\ref{self2}) and (\ref{self1}). 
We therefore computed them numerically by ${\bf k}$-summation on reciprocal 
lattices of sizes up to $8000\times8000$.
Introducing the notation 
\begin{equation}
\gamma := \frac{U\Delta}{2}  
\end{equation}
for the energy gap of the Hartree dispersion (\ref{HFenergies}),
we notice that both the integrals in (\ref{self2}) and (\ref{self1})
only depend on $\gamma$. 

\begin{figure}[t]
\begin{center}
\epsfig{file=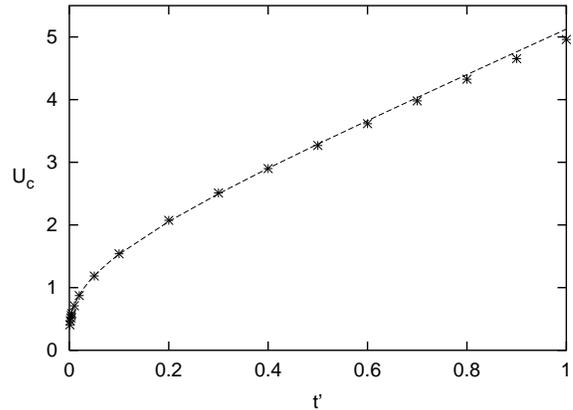,width=0.93\linewidth}
\end{center}
\caption{\label{fig:UC_2d}Critical interaction $U_c$ vs. next-nearest neighbor hopping 
matrixelement $t'$ in $d=2$. Dashed line: asymptotic behaviour, (\ref{approx2d}).}
\end{figure}

We find that for any $t'\ne 0$, the integral in (\ref{self1})
has a finite limit for $\gamma\to0$. This implies that below a critical 
value $U_c$ of the interaction a stable solution with $\Delta>0$ no longer exists. 
For $d=2$ this fact was already noted  by Lin and Hirsch \cite{Lin 87}.
The dependence of $U_c$ on $t'$ is shown in fig.\ \ref{fig:UC_2d}. 
\noindent For $t'\to 0$, $U_c(t')$ decreases very slowly. 
An analytic description of the behavior of 
$U_c(t')$ may be obtained by noting that the divergence of the integral in (\ref{self1})
for $t'=0$  is cut off just at the value of $\gamma$ where the 
indirect bandgap \cite{bandgap}
\begin{figure}[t]
\begin{center}
\epsfig{file=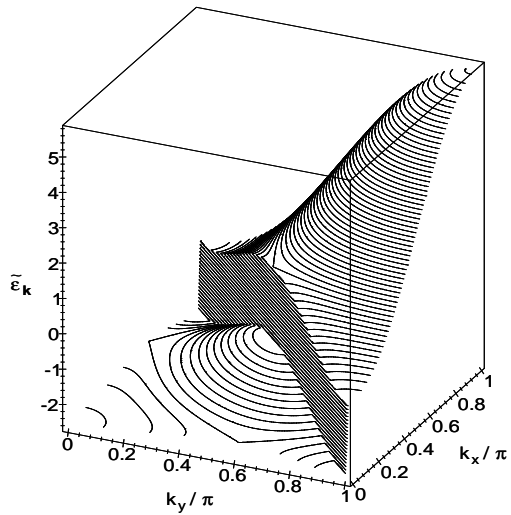,
width=0.9\linewidth,height =0.9\linewidth,angle=0,bbllx=0,bblly=40,
bburx=440,bbury=555}
\end{center}
\caption{\label{fig:hartree_dispersion}Hartree dispersion 
$\tilde{\epsilon}_{\bf k}$ in $d=2$ for $t'=0.4$ and 
$\gamma=1.0$ with an indirect bandgap between $(\pi/2, \pi/2)$ and $(0,\pi)$.}
\end{figure}
\noindent in the Hartree dispersion (\ref{HFenergies}) 
closes, namely at 
\begin{equation} \label{close_bandgap}
\gamma^* =  2t'.
\end{equation}
The asymptotic behaviour of $U_c(t')$ for small $t'$ may thus be found by 
setting $t'=0$ in (\ref{self1}) and using $\gamma^*$ as a cut-off. This leads to
\begin{equation}  \label{approx2d}
[U_c(t')]^{-1} =  c_1 \int_0^4 d\epsilon \, K\left(\sqrt{1 - 
\left(\frac{\epsilon}{4}\right)^2 } \right) \frac{1}{\sqrt{\epsilon^2 + (2t')^2}}
\end{equation}
where $K$ is the complete elliptic integral and $c_1 = 1/(2 \pi^2)$.
The approximation (\ref{approx2d}) is very 
good up to $t'\approx 1$, as can be seen from fig.\ \ref{fig:UC_2d}.
For small $t'$, (\ref{approx2d}) reduces to \cite{van Dongen statement}  
\begin{equation}
[U_c(t')]^{-1} = \frac{c_1}{2}\, (\ln t')^2 + c_2\,\ln t'
\end{equation}
where $c_2\approx -0.15$. 
The quadratic dependence of $U_c$ on $\ln t'$ is characteristic for $d=2$, where
the logarithmic divergence of the density of states (DOS) for $t'=0$ at the 
Fermi level introduces an additional factor $\ln t'$. 

The behaviour of the order parameter $\Delta$ at $U\approx U_c$ can be characterized
as follows:
For $t'$ below a threshold value $t_0' \approx 0.4$ the order parameter $\Delta$ 
jumps to $0$ discontinuously when the interaction 
is reduced below $U_c(t')$, indicating a first order phase transition.
For $U>U_c$, the order parameter takes exactly 
the same value as for $t' = 0$.
For $t'>t_0'$ and $U\to U_c$ the order parameter goes to zero continuously, 
but extremely fast. 

\subsection{$d=3$}
\begin{figure}[t]
\begin{center}
\epsfig{file=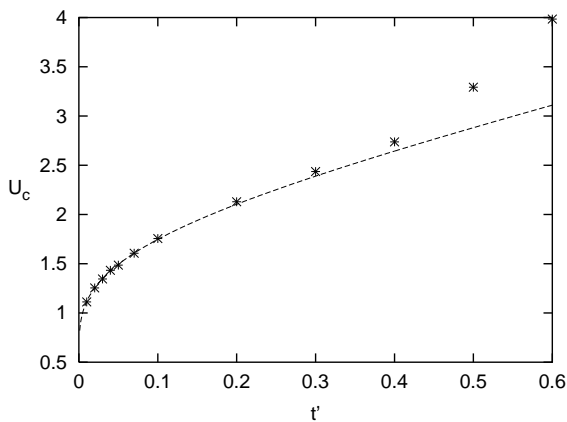,width=0.93\linewidth}
\end{center}
\caption{\label{fig:UC_3d} Critical interaction $U_c$ vs.\ $t'$ in $d=3$.
Dashed line: asymptotic behavior, (\ref{logarithmic}).}
\end{figure}

We evaluate the Hartree equations numerically
at zero temperature, using lattice sizes up to $400\times 400\times 400$.
As before we notice that the divergence of the integral in (\ref{self1}) 
for $\gamma\to0$
is cut off for any finite $t'$. The resulting critical interaction $U_c(t')$ is
shown in fig.\ \ref{fig:UC_3d}.
The indirect bandgap again closes at $\gamma^* = 2 t'$,
but in this case the order parameter goes to zero continuously as $U\to U_c$,
indicating a second-order phase transition. 
The logarithmic divergence of the integral in (\ref{self1}) is cut off
at $\gamma^*$, which leads to the asymptotic expression \cite{van Dongen statement} 
\begin{equation} \label{logarithmic}
[U_c(t')]^{-1} = -c_3 \ln t' + c_4,
\end{equation}
where $c_4 \approx 0.25$ is a fit parameter and  
$c_3 = N(0)\approx 0.14$, with $N(0)$ as the DOS at the Fermi level in $d=3$
for $t'=0$. 
Eq.\ (\ref{logarithmic}) represents a good fit for $t'<0.4$. 

\subsection{$d=\infty$}
\begin{figure}[t]
\begin{center}
\epsfig{file=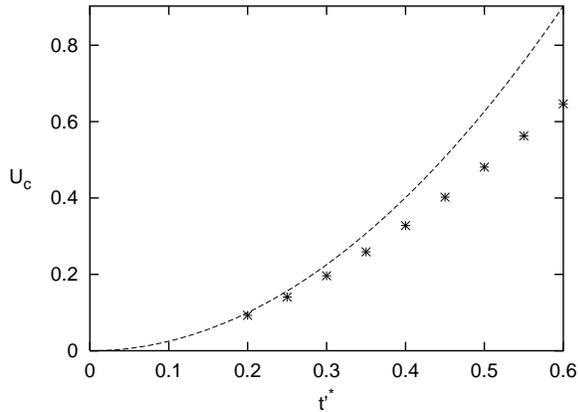,width=0.93\linewidth}
\end{center}
\caption{\label{fig:UC_dinf} Critical interaction versus $t'^*$ in 
$d=\infty$. Dashed line: asymptotic behavior, (\ref{dinf_asympt}).}
\end{figure}

To analyze the model (\ref{Hamiltonian})
on a hypercubic lattice in the limit of infinite dimensions,  one 
has to ensure that the hopping remains finite; this is achieved by the scaling 
\cite{Metzner 89,Mueller-Hartmann 89}
\begin{equation} \label{scaling}
t = \frac{t^*}{\sqrt{2d}}\ ,\ \ \ \ \ \ \  t' = \frac{{t'}^*}{\sqrt{2d(d-1)}} 
\end{equation}
for the NN- and NNN-terms, respectively. In the following 
we set $t^* = 1$.
As shown by M\"uller-Hartmann \cite{Mueller-Hartmann 89}  the dispersion 
of the free electrons then reduces to
\begin{equation}
\epsilon({\bf k}) \to  \epsilon - \frac{t'^*}{\sqrt{2}} (1-\epsilon^2).
\end{equation}
Here $\epsilon$ is the kinetic energy without the $t'^*$-term, which has 
a Gaussian distribution \cite{Metzner 89}
\begin{equation}
N(\epsilon) = \frac{1}{\sqrt{2\pi}} \exp\left(\frac{\epsilon^2}{2}\right).
\end{equation}
As a result,  we can write the Hartree dispersion in the form
\begin{equation} \label{HFdinf}
\tilde{\epsilon} = \frac{1}{2} \left\{+\sqrt{2}\, t'^* \left(\epsilon^2 - 1\right)
+ 2 \epsilon \sqrt{1 + \frac{U^2 \Delta^2}{4 \epsilon^2} } \right\}; 
\end{equation}
as before, it only depends on $t'^*$ and  $\gamma = U\Delta /2$.
We note that due to the exponential band-tails of the hypercubic DOS there is 
no true gap between the two Hartree bands for any value of $\gamma$. 
The Hartree equations can be written as one-dimensional integrals and are readily solved 
numerically. Again we find a critical interaction $U_c$ for the stability of
long-range antiferromagnetic order at $T=0$. A plot of $U_c({t'}^*)$ is shown in fig.\ 
\ref{fig:UC_dinf}.
Due to the absence of a band gap, the asymptotic behavior of $U_c$ for small $t'^*$
is qualitatively different from that in $d=2,3$. 
In the present case, the cut-off for the integral in 
(\ref{self1}) is provided by a small filling in the upper one of the two Hartree 
bands (\ref{HFdinf}) which vanishes exponentially for $t'\to 0$ i.e.\
$\gamma^* \approx \frac{|t'^{*}|}{2}\, \exp(-\frac{1}{(t'^{*})^2})$. 
This leads to an asymptotic behaviour \cite{van Dongen statement} of the form
\begin{equation} \label{dinf_asympt}
U_c = c\, (t'^*)^2
\end{equation}
where $c\approx \sqrt{2\pi}$. Hence $U_c$ decreases much faster than in $d=2,3$.
As in $d=3$ the order parameter $\Delta$  goes to zero continuously for $U \to U_c$.
A second-order phase transition is therefore seen to be generic for $d>2$.
We note that in $d=3$ , the continuously vanishing order parameter
implies a (small) range of $U$ values where the band gap is zero but $\Delta$ 
is still finite. This implies the existence of a \emph{metallic} 
antiferromagnetic phase for $t'\ne 0$ as shown in fig.\ 5.  
\begin{figure}[t]
\begin{center}
\epsfig{file=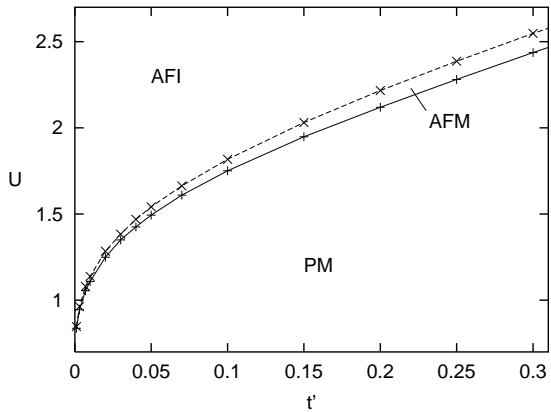, width=0.93\linewidth}
\end{center}
\caption{\label{fig:metallic_AF} Region of stability of a metallic antiferromagnetic 
phase (AFM) in $d=3$ in the $U$-$t'$ phase diagram. Full line: $U_c(t')$, dashed line: 
value of $U$ where the band gap closes. Outside the AFM region the system is either
an antiferromagnetic insulator (AFM) or a paramagnetic metal (PM).}
\end{figure}
In $d=\infty$, on the other hand, there is never a true bandgap anyway, 
as described above. Therefore, for $t'\ne 0$ and $U>U_c$ 
we always find metallic antiferromagnetism.
The existence of such a phase was already discussed by 
Georges et al.\ \cite{Georges 96}. Quite generally, in dimensions $2\le d < \infty$ 
a next-neighbour hopping term $t'$ is found to be very effective in suppressing the
perfect nesting instability towards an antiferromagnetic ground state. 

\subsection*{Acknowledgments}We would like to thank P.G.J.\ van\ Dongen,
A.\ Chubukov, K.\ Held, A.\ Kampf, W.\ Metzner, M.\ Ulmke and, in particular, 
J.\ Schlipf 
for several useful discussions.

\subsection*{Note added in proof:}
\noindent In $d=2$, if the $t'$-hopping occurs only along \emph{one} of the diagonals
of the squares, the dispersion reads
\mbox{$\epsilon_{\bf k} = -2 t \, (\cos k_x +\cos k_y ) + 2 t' \cos(k_x + k_y)$}.
The corresponding $t-t'$-Hubbard model may be viewed as a simple dimer model
of the two-dimensional organic superconductor $\kappa - ({\rm BEDT-TTF})_2 {\rm X}$
which was investigated by Kino and Fukuyama [J.\ Phys.\ Soc.\ Jpn.\ {\bf 65} 
(1996) 2158] within Hartree-Fock theory. In contrast to the results
of section \ref{d_equals_two}, one then obtains a region of stability of an AFM phase 
which is qualitatively similar to the result for the regular $t-t'$-model in 
$d=3$ shown in \mbox{fig.\ 5.} 
The discontinuous transition from the paramagnetic 
metal into the antiferromagnetic insulator obtained by Kino and Fukuyama
is found to occur for $|t'| > 0.6$ when this AFM phase is unstable. \\
We thank R.\ McKenzie for drawing our attention to this variant of the 
$t-t'$-Hubbard model, and to H.\ Kino and H.\ Fukuyama for correspondence.

\newpage


\begin{thebibliography}{99}
\bibitem{bandgap} By this we mean a bandgap 
with nonzero wavevector difference between the valence band maximum 
and the conduction band minimum, see fig. \ref{fig:hartree_dispersion}.
\end{thebibliography}
\end{document}